\documentclass[11pt]{article}
\usepackage[utf8]{inputenc}
\usepackage[english]{babel}
\usepackage{amssymb}
\usepackage{docmute}
\usepackage{booktabs}
\usepackage{authblk}
\usepackage{algorithm}
\usepackage{algpseudocode}
\usepackage{titletoc}
\usepackage{graphicx}
\usepackage{amsmath,amssymb,amsthm,bm}
\usepackage{pdfpages}
\usepackage{subcaption}
\usepackage{listings}
\usepackage{latexsym,color,minipage-marginpar,caption,multirow,verbatim}
\usepackage{enumerate}
\usepackage{xcolor}
\usepackage{nameref}
\usepackage{mathtools}
\usepackage[sort]{natbib}
\bibpunct[, ]{[}{]}{,}{}{}{,}
\setcitestyle{numbers}
\usepackage{hyperref}
\usepackage{cleveref}
\hypersetup{
  colorlinks=true,
  citecolor=blue,
  linkcolor=blue,
  urlcolor=blue}

\makeatletter
\let\start@align@nopar\start@align
\let\start@gather@nopar\start@gather
\let\start@multline@nopar\start@multline
\long\def\start@align{\par\start@align@nopar}
\long\def\start@gather{\par\start@gather@nopar}
\long\def\start@multline{\par\start@multline@nopar}
\makeatother

\newcommand{\fref}[1]{Figure~\ref{#1}}

\usepackage{xr}

\makeatletter
\newcommand*{\addFileDependency}[1]{
\typeout{(#1)}
\@addtofilelist{#1} 
\IfFileExists{#1}{}{\typeout{No file #1.}}
}\makeatother

\def\subsubsection#1{\bigskip\noindent\textbf{#1}.}

\usepackage[top=1in,bottom=1in,left=1in,right=1in]{geometry}

\title{Tree Thinking in the Genomic Era: Unifying Models Across Cells, Populations, and Species}
\author[1, *]{Yun Deng}
\author[2, 3, 9]{Shing H. Zhan}
\author[4, 9]{Yulin Zhang}
\author[5, 6, *]{Chao Zhang}
\author[7, 8, *]{Bingjie Chen}

\affil[1]{\footnotesize Department of Genetics, Stanford University, Stanford, CA 94305, United States of America}
\affil[2]{\footnotesize Big Data Institute, Li Ka Shing Centre for Health Information and Discovery, University of Oxford, United Kingdom}
\affil[3]{\footnotesize Infectious Disease Epidemiology Unit (IDEU), Nuffield Department of Population Health, University of Oxford, Oxford, United Kingdom}
\affil[4]{\footnotesize Center for Computational Biology, University of California, Berkeley, CA 94720, United States of America}
\affil[5]{\footnotesize School of Life Sciences, Peking University, Beijing 100871, China}
\affil[6]{\footnotesize Peking-Tsinghua Center for Life Sciences, Peking University, Beijing 100871, China}
\affil[7]{\footnotesize GMU-GIBH Joint School of Life Sciences, The Guangdong-Hong Kong-Macao Joint Laboratory for Cell Fate Regulation and Diseases, Guangzhou Laboratory, Guangzhou Medical University, Guangzhou 510260, China}
\affil[8]{\footnotesize State Key Laboratory of Respiratory Disease, Department of Critical Care, Second Affiliated Hospital, Guangzhou Medical University, Guangzhou 510260, China}
\affil[9]{\footnotesize These authors contribute equally as co-second authors.}
\affil[*]{\footnotesize Corresponding authors: 
Yun Deng (yundeng@stanford.edu), 
Chao Zhang (chaozhang@pku.edu.cn), 
Bingjie Chen (bingjiechen@gzhmu.edu.cn)}
\date{}

\begin{document}

\maketitle

\abstract{
The ongoing explosion of genome sequence data is transforming how we reconstruct and understand the histories of biological systems. Across biological scales--from individual cells to populations and species--trees-based models provide a common framework for representing ancestry. Once limited to species phylogenetics, “tree thinking” now extends deeply to population genomics and cell biology, revealing the genealogical structure of genetic and phenotypic variation within and across organisms. Recently, there have been great methodological and computational advances on tree-based methods, including methods for inferring ancestral recombination graphs in populations, phylogenetic frameworks for comparative genomics, and lineage-tracing techniques in developmental and cancer biology. Despite differences in data types and biological contexts, these approaches share core statistical and algorithmic challenges: efficiently inferring branching histories from genomic information, integrating temporal and spatial signals, and connecting genealogical structures to evolutionary and functional processes. Recognizing these shared foundations opens opportunities for cross-fertilization between fields that are traditionally studied in isolation. By examining how tree-based methods are applied across cellular, population, and species scales, we identify the conceptual parallels that unite them and the distinct challenges that each domain presents. These comparisons offer new perspectives that can inform algorithmic innovations and lead to more powerful inference strategies across the full spectrum of biological systems.}

\section*{Significance}
Tree-based models lie at the heart of evolutionary biology, yet they have traditionally been developed within separate disciplines--phylogenetics for species, coalescent theory for populations, and lineage tracing for cells. By highlighting conceptual and methodological parallels across these fields, we show how similar algorithmic and statistical challenges recur at different biological scales. Recognizing these shared principles facilitates the transfer of ideas--such as efficient algorithms, probabilistic inference, and model interpretation.

\section{Introduction}

Trees are the simplest form of graph data structures in computer science, yet they allow a remarkably expressive abstraction for evolutionary history in biology. This conceptual link dates back to Charles Darwin, who famously sketched the first evolutionary tree in his Notebook B as a branching diagram \citep{de1960darwin}. Darwin’s ``Tree of Life" was more than a metaphor--it represents a profound shift in understanding life’s diversity as the result of common descent and the speciation process \citep{goldstein2009charles}.

Over the next one and a half centuries, the tree has evolved into a central principle for describing the relationships of lineages at every scale of biological organization. From the divergences among species to the genealogies of the individuals in a population and the developmental histories of cells, trees are foundational for representing and interpreting evolutionary processes \citep{zou2024common, brandt2024promise, mao2025cell}.

\begin{figure}
    \centering
    \includegraphics[width=\linewidth]{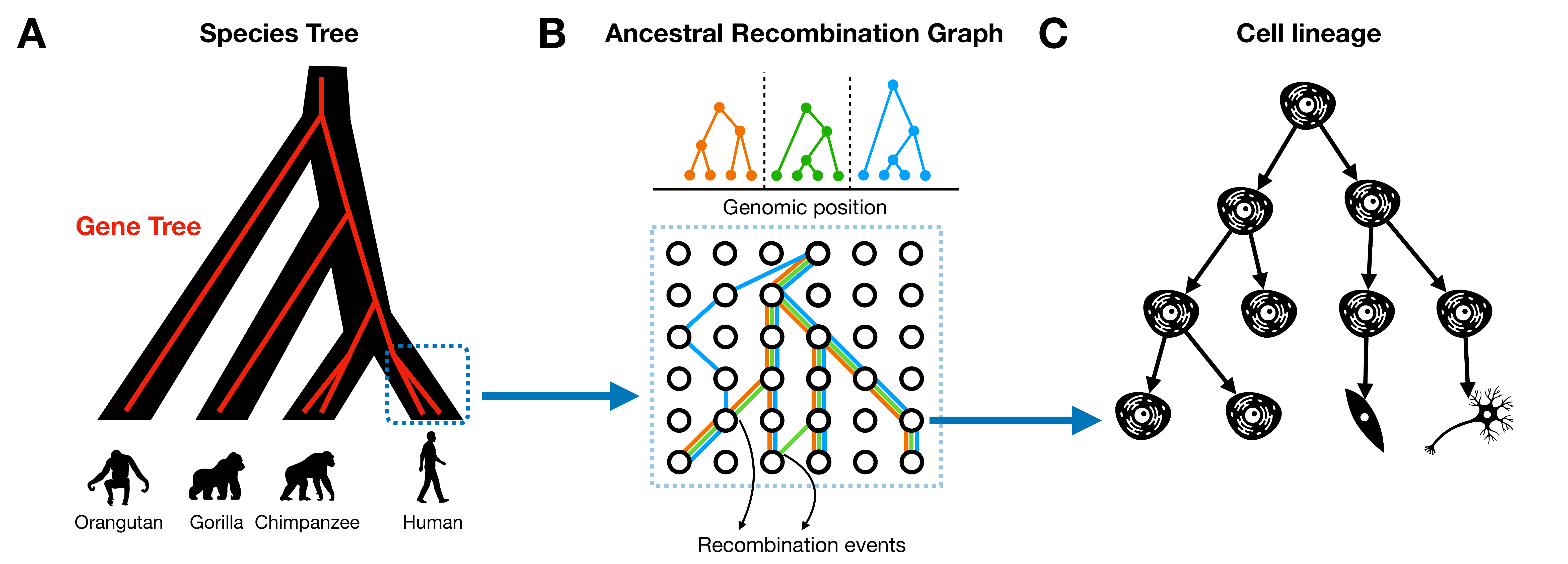}
    \caption{
    Tree structures across biological scales.
    (A) A species tree at the macroevolutionary level showing relationships among human, chimpanzee, gorilla, and orangutan. An embedded gene tree (pink) illustrates the genealogical history at a specific locus.
    (B) An Ancestral Recombination Graph (ARG) at the population level, representing the genealogical history of three genomic loci (red, cyan, and yellow) separated by two recombination events. The ARG can also be represented by local trees along the genome that correspond to different non-recombining blocks.
    (C) A cell lineage tree at the individual level depicting relationships among cells generated through cell division and differentiation.
    }
    \label{fig:fig1}
\end{figure}

The species tree sits at the top of this hierarchy, summarizing the broad strokes of evolution: the series of divergence and speciation events that have generated the tapestry of life \citep{swenson2012gene, szollHosi2015inference}. In phylogenetics, the species tree represents an idealized history of how populations split and evolve into distinct species. The ambition to reconstruct the complete “Tree of Life” has driven decades of research, drawing on morphological, molecular, and genomic data to infer the branches of the speciation process \citep{schliep2011phangorn, tamura2013mega6, yang2007paml} (\fref{fig:fig1}A).

A more complex reality is embedded within the species tree: Each gene in the genome has its own genealogical history, known as gene trees \citep{swenson2012gene} (\fref{fig:fig1}A). The species tree is best understood as a statistical abstraction--a composite or a consensus of the underlying set of gene trees along the genome. However, this gene tree–centric view is incomplete. While each locus may have a well-defined genealogy, recombination is the reason that these genealogies vary along the genome. A more general structure is the Ancestral Recombination Graph (ARG), which captures how local genealogies change and how they are correlated by recombination events across the genome \citep{lewanski2023era, nielsen2024inference, brandt2024promise} (\fref{fig:fig1}B).

The concept of an ARG is first studied at the population level through the coalescent with recombination \citep{griffiths1981neutral, hudson1983properties}. For any single locus, the ancestral relationships between the sampled individuals can be represented by a coalescent tree--by tracing their ancestors in the previous generations through a stochastic process called the coalescent. However, when a recombination event is encountered, the ancestral lineages which contribute to the recombination event must be tracked. The resulting structure is not a single tree, but an ARG, which is a network-like structure (\fref{fig:fig1}B). This graph structure can be interpreted as a sequence of gene trees on corresponding non-recombining blocks, with recombination describing how adjacent coalescent trees change from one to another (\fref{fig:fig1}B).

At a finer scale, tree structures also arise in the context of cell lineages. During development, cells divide and differentiate clonally, giving rise to a hierarchical tree of descent from the zygote to all the cells in the body (\fref{fig:fig1}C). Cell lineages do not undergo recombination, with each round of cell division producing two progeny cells that inherit the genetic material of their parent. In this sense, cell lineage trees are a complementary, non-recombining analogue to population genealogies--extending the utility of tree-based thinking into developmental and cancer biology \citep{yang2022lineage, nathans2024genetic}.

Although these tiers of evolution differ in terms of rates and mechanisms--macroevolution by speciation, microevolution by drift and gene flow, and ultra-micro cellular evolution \cite{wu2016ecology} via somatic mutation and clonal selection--the underlying logic of tree-based modeling unites them. Tree thinking thus provides a shared language for understanding descent, diversification, and the processes that generate biological diversity across scales.

\section{Connections between tree-based methods across different domains}

Although developed to study different biological systems, tree-based methods used in phylogenetics, population genetics, and cell lineage tracing share common principles. They aim to infer branching histories from genetic sequence data and to link genealogical structures to the underlying evolutionary or functional processes. Recognizing these biological and computational parallels reveals conceptual unity across scales and allows ideas and algorithms developed in one domain to inspire advances in another. Below we illustrate some deep connections using a few examples from different domains.

\begin{figure}
    \centering
    \includegraphics[width=\linewidth]{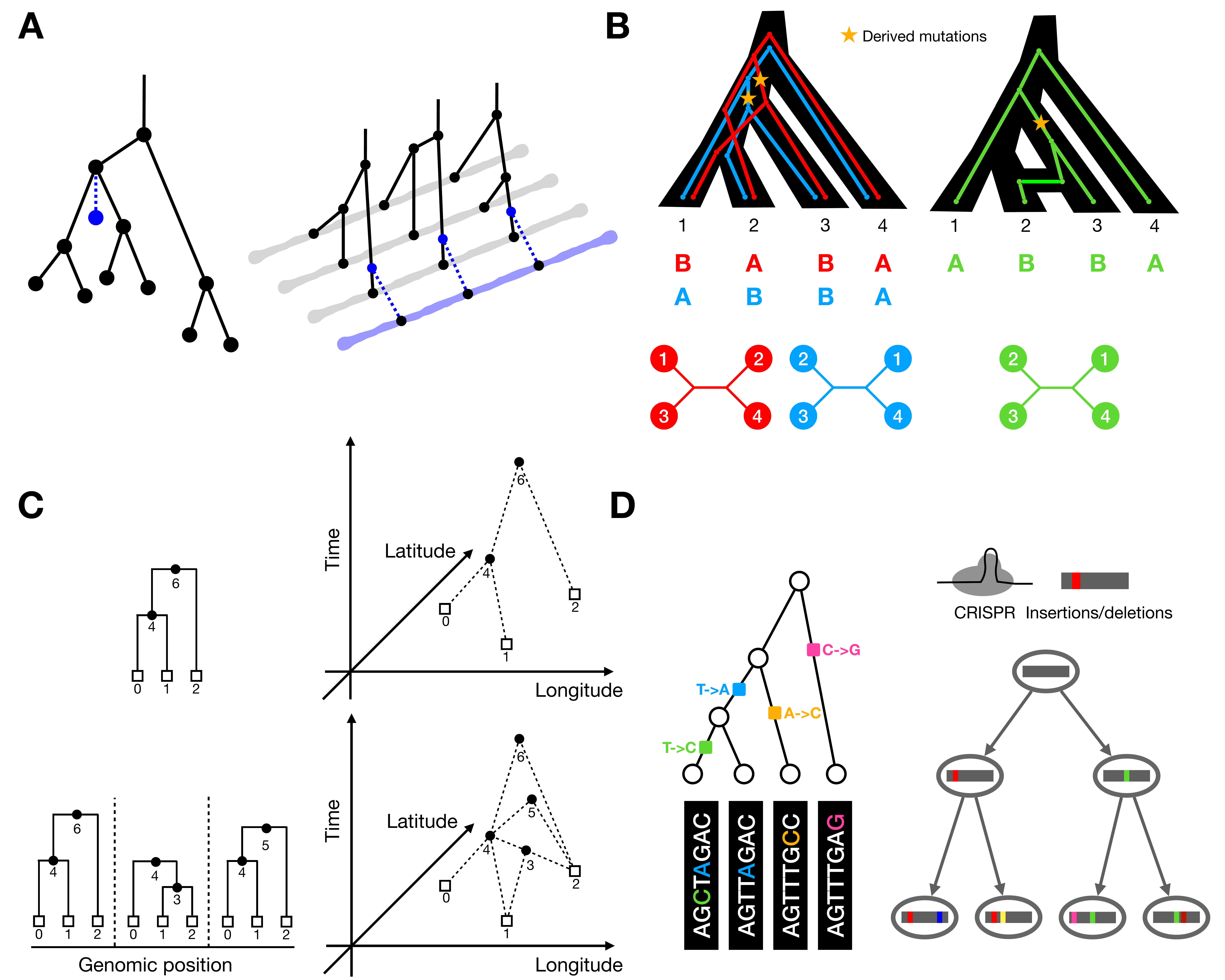}
    \caption{
    Conceptual links between tree-based methods across biological domains.
    (A) Sample addition. In phylogenetics, phylogenetic placement adds a new sampled sequence (blue) onto an existing tree (left). Similarly, in ARG reconstruction, threading finds the joining points of a new haplotype with respect to each local tree along the genome (right).
    (B) Inference of reticulate relationships. Both the D-statistic (ABBA–BABA test) in population genetics and quartet-based methods in phylogenetics detect gene flow by quantifying asymmetry in tree topologies which arise from excess shared ancestry between species or populations connected by genetic exchange.
    (C) Migration reconstruction. Phylodynamic inference estimates the geographic locations of ancestral nodes over time in a single tree (top), while ARG-based methods infer the geographical locations of all the ancestral nodes in all the marginal trees (bottom).
    (D) Genealogy inference. In phylogenetics and population genetics, gene trees are typically inferred from nucleotide or amino acid differences observed in multiple sequence alignments (left), whereas cell lineage trees in single-cell studies are reconstructed from CRISPR-induced insertions and deletions that serve as mutational barcodes (right).
    }
    \label{fig:fig2}
\end{figure}

\paragraph{\textbf{ARG inference}}
Methods to infer ARGs have drawn heavily from developments in phylogenetics, as the ARG can be viewed as a sequence of correlated gene trees. Classical tree-building algorithms, such as UPGMA and Neighbor Joining \citep{felsenstein2004inferring}, demonstrate how pairwise distance matrices can be used to construct tree topologies--a principle adopted by modern ARG inference tools. For instance, Relate \citep{Speidel2019} uses a modified Li-Stephens model \citep{Li2003} to compute local pairwise distance matrices and then infer local genealogical trees, with recombination breakpoints called with transitions of topology. Although tailored for population-genetic assumptions, the core logic of Relate parallels distance-based phylogenetic reconstruction. Another foundational operation, threading, determines how a new genetic sequence integrates into an existing ARG with respect to all local trees (\fref{fig:fig2}A). First implemented in ARGweaver \citep{Rasmussen2014c} and later extended by ARG-Needle \citep{zhang2023biobank} and SINGER \citep{deng2025robust}, threading generalizes the classic phylogenetic placement problem \citep{matsen2010pplacer} from single trees to ARGs.

\paragraph{\textbf{Real-time ARG inference for SARS-CoV-2}}
During the COVID-19 pandemic, SARS-CoV-2 genomes were sequenced at unprecedented speed and scale, with GISAID \citep{GISAID_2017} containing over 17.5 million SARS-CoV-2 genomes as of October, 2025. Although this data deluge overwhelms classical phylogenetic methods and tools (e.g., FastTree \citep{FastTree_2009} and IQTree2 \citep{IQTree2_2020}), it has motivated the development of new phylogenetic methods and tools, notably UShER \citep{UShER_2021}, which can handle modern outbreak-scale sequence data in real time as genomes are collected. UShER achieves scalability by exploiting a characteristic of the densely sampled COVID-19 dataset: most sequences are identical or nearly identical to their closest matching sequences \citep{Ye2022,Kramer2023}. This means that the sequences can be highly compressed for computational speedups and that they are amenable for parsimony-based inference, which is far less computationally costly than likelihood-based inferences. However, UShER's underlying data model represents evolutionary history as a single tree, which does not allow recombination. A separate parsimony-based method has been developed to detect recombination \emph{post hoc} by looking for long branches in an existing UShER tree \citep{RIPPLES_2022}. More recently, sc2ts has been developed to build large SARS-CoV-2 genealogies in the form of ARGs \citep{Sc2ts2_2025}, integrating recombination detection into genealogical inference in a single cohesive framework. Sc2ts combines the compact tree sequence format \citep{tskitKelleher2018}, a highly efficient Hidden Markov Model inference engine \citep{Kelleher2019}, and parsimony-based heuristics to enable real-time reconstruction of an ARG containing $\sim$2.48 million SARS-CoV-2 genomes, which represents the pandemic phase of SARS-CoV-2 evolutionary history.

\paragraph{\textbf{Gene flow detection}}
Phylogenetic and population genetic approaches to detect gene flow are both grounded in identifying asymmetries in genealogical relationships that violate a strictly tree-like model of evolution. In phylogenetics, reticulate evolution--such as hybridization or introgression--is inferred when the frequencies of alternative four-taxon (quartet) topologies become imbalanced, as these asymmetries indicate excess coalescence between taxa exchanging genes (\fref{fig:fig2}B). Quartet-based network methods \citep{solis2017phylonetworks, wen2018inferring} formalize this intuition by quantifying deviations from the expected distribution of topologies under the multi-species coalescent. In population genetics, the ABBA–BABA statistic \citep{Green2010AGenome} applies the same logic at the level of site patterns rather than inferred trees. By comparing the counts of biallelic configurations consistent with alternative genealogies, it measures the degree of allele sharing between non-sister populations beyond what incomplete lineage sorting alone would produce. Thus, both the frameworks--quartet imbalance in phylogenetics and ABBA–BABA statistics in population genetics--capture the same underlying signal of gene flow: asymmetric genealogical relationships induced by hybridization or introgression (\fref{fig:fig2}B).

\paragraph{\textbf{Migration inference}}
Recovering the evolutionary and dispersal processes that shape the spatial distribution of populations or taxa--a central aim of phylogeography--is a key application of phylogenetics in epidemiology \citep{kuhnert2011phylogenetic}. Classical comparative methods, such as Phylogenetic Generalized Least Squares, are widely applied to infer spatial diffusion by treating geographic coordinates as traits and using regression models to estimate their ancestral states \citep{martins2004compare, biek2006virus}. ARG-based methods for inferring population migration histories represent a natural extension of these tree-based phylogeographic approaches (\fref{fig:fig2}C). Whereas classical methods rely on a single tree \citep{avise2000phylogeography, knowles2009statistical, knowles2002statistical}, ARG-based frameworks, such as GAIA \citep{grundler2025geographic}, tsdate \citep{wohns2022unified}, and spacetrees \citep{osmond2024estimating}, leverage the full sequence of locally correlated genealogies along the genome. For example, GAIA estimates ancestral locations by applying a minimum-migration-cost function to ancestral haplotypes in a generalized parsimony framework--a parametric extension of the maximum parsimony methods commonly used in phylogeography \citep{slatkin1989cladistic, swofford2003paup, maddison2005macclade, wallace2007statistical}. Similarly, methods like spacetrees model sample locations as continuous variables using diffusion processes (e.g., branching Brownian motion), building directly upon comparative method frameworks from phylogeography \citep{felsenstein1985phylogenies, harvey1991comparative}.

\paragraph{\textbf{Cell lineage}}
Cell lineage research, focused on decoding cellular ancestry via somatic mutations or genomic barcoding (\fref{fig:fig2}D), shares core methodologies with phylogenetics and population genetics. First, marker-based tree reconstruction is conceptually shared across domains: just as phylogenetics uses nucleotide or amino acid substitutions, cell lineage tracing uses endogenous genomic alterations (e.g., copy number variants in tumor cells \citep{nik2012life}, mitochondrial mutations \citep{ludwig2019lineage}) or synthetic barcodes, such as CRISPR-edited GESTALT arrays \citep{mckenna2016whole}. Distance-based algorithms, including neighbor joining, are similarly adapted to calculate ``clonal distance" between cells. Second, conflicts between local and global trees mirror the gene tree–species tree problem. Also, probabilistic modeling of uncertainty is essential across all scales. Single-cell sequencing errors and barcode dropout \citep{salvador2019possible} have led to Bayesian and probabilistic approaches--such as Waddington Optimal Transport \citep{schiebinger2019optimal}--that parallel Bayesian inference frameworks for phylogenies and ARGs.

\section{Ongoing challenges}

Despite rapid methodological progress, the inference and analysis of tree structures across biological scales still face substantial challenges.

\paragraph{\textbf{Phylogenetic inference}}
Phylogenetic reconstruction under incomplete lineage sorting and recombination presents a persistent dilemma. Recombination-free “coalescent genes” are typically too short to produce well-resolved local trees, whereas fixed-length windows may span multiple recombination events, distorting true genealogical relationships. Methods based on site pattern frequencies avoid explicit phylogenetic tree reconstruction but can be biased under differential evolutionary rates across taxa. Inferring ARGs offers a principled solution, but modeling recurrent mutations under a finite-sites model, especially over deep timescales, remains computationally and statistically challenging. Addressing these issues will require new probabilistic frameworks that balance biological realism, scalability, and interpretability across evolutionary depths.

\paragraph{\textbf{ARG inference}}
Current ARG inference algorithms remain constrained by simplifying assumptions about evolutionary models. Many are derived under the assumptions of neutrality, panmixia, and constant population size, which poorly reflect biological systems shaped by pervasive background selection, population structure, and complex demography. Widely used frameworks--such as those based on the Li–Stephens model--are optimized for large, high-quality human datasets and often perform sub-optimally for non-model species or limited sample sizes. Another challenge lies in quantifying uncertainty: ARG estimation based on short non-recombining segments is inherently noisy, and point estimates may obscure this ambiguity. Robust uncertainty propagation is essential for reliable downstream analyses, particularly for fine-scale applications such as selection detection or local ancestry inference, where overconfident estimates can yield misleading conclusions. Finally, downstream ARG analyses remain heavily focused on summary statistics, providing only incremental gains in accuracy. A major step forward will be to develop novel applications that directly leverage genealogical features to address questions inaccessible to classical summary-statistic frameworks.

\paragraph{\textbf{Cell lineage tracing}}
Recent advances in single-cell genomics and lineage-tracing technologies have established tree-based models as central for studying cell developmental history. Two complementary strategies dominate: somatic mutation–based tracing, which uses naturally occurring variants as clonal markers, and synthetic barcode tracing, which introduces heritable edits via CRISPR systems. These methods enable reconstruction of developmental hierarchies, such as germ layer specification, and track tumor evolution, revealing clonal expansion and resistance to therapy. Yet, lineage tree inference faces three major challenges. First, computational scalability: datasets encompassing millions of cells render current distance-based and heuristic methods intractable. Second, uncertainty quantification: barcode dropout and false positive mutations complicate probabilistic modeling, and lineage tracing still lacks standardized noise frameworks analogous to those in phylogenetics or population genetics. Third, biological integration: reconstructed trees capture ancestry but often remain disconnected from transcriptomic and spatial information, limiting biological interpretation. A unified framework linking developmental history, cell state, and spatial position remains a key frontier.

\section{Conclusion and future perspectives} 

Tree-based models have become indispensable for describing biological processes across scales, from species evolution to cellular development. Looking ahead, several research directions stand out as particularly promising for advancing the scope and impact of tree-based and ARG-based methodologies.

\begin{figure}
    \centering
    \includegraphics[width=\linewidth]{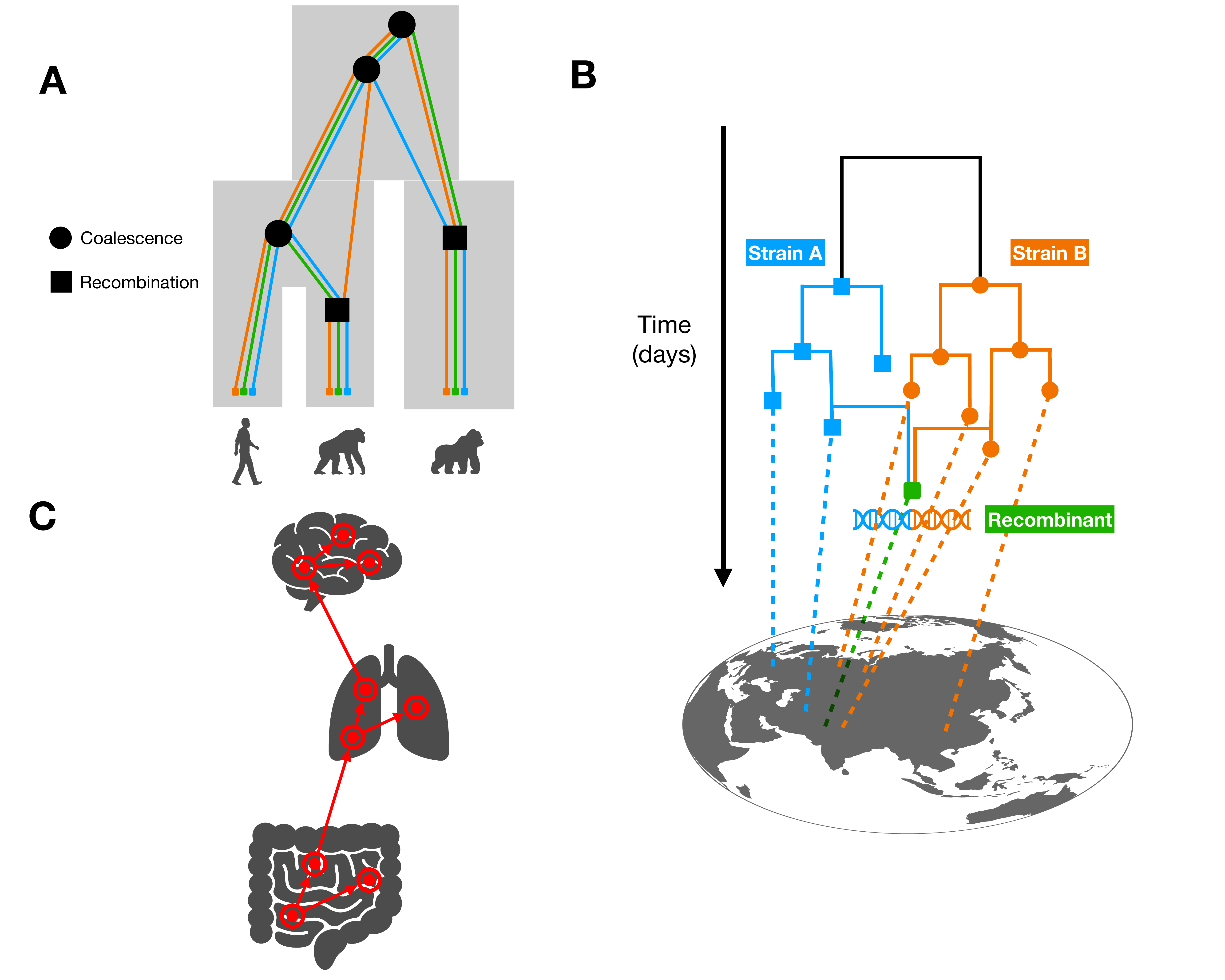}
    \caption{
    Future directions for tree-based methods across biological domains.
    (A) An Ancestral recombination graph (ARG) embedded in a species tree, illustrating the full network of coalescence events and recombination events in the multiple species coalescent.
    (B) A real-time pathogen ARG integrating longitudinal genome sequence data and geographic locations, enabling joint inference of genealogical relationships, mutation, and recombination over time.
    (C) Tumor cell lineage tracing annotated with organ and tissue locations, providing a framework to reconstruct metastatic trajectories and to infer patterns of cellular migration across the body.
    }
    \label{fig:fig3}
\end{figure}

\paragraph{Multi-species ARGs}
In phylogenetics, there is a growing shift from representing evolution as a single global species tree toward modeling the distribution of local gene trees along the genome \citep{burbrink2025recombination} (\fref{fig:fig3}A). While the multi-species ARG naturally captures how genealogies vary across loci, most current phylogenetic frameworks still rely on reconstructing trees for predefined regions (such as genes). Extending ARG methodologies to multi-species contexts could unify gene tree inference and comparative genomics. Recent developments point in this direction. TRAILS \citep{rivas2024trails} models a four-taxa ARG across species using a coalescent HMM framework \citep{hobolth2007genomic}, while SINGER \citep{deng2025robust} successfully infers an ARG for the highly divergent HLA region in humans, recovering patterns consistent with trans-species polymorphism among primates \citep{fortier2022ancient} from population-level genetic data.

\paragraph{Pathogen ARGs}
The COVID-19 pandemic has highlighted a striking imbalance between our sequencing capacity and analytical scalability. While new methods such as UShER and sc2ts have demonstrated that pandemic-scale reconstruction of phylogenies and ARGs is feasible, key challenges remain (\fref{fig:fig3}B). Future work should prioritize: (1) scalable inference of trees and ARGs for large but less densely sampled genomic datasets compared to SARS-CoV-2; (2) Bayesian approaches to capture uncertainty in outbreak-scale ARGs, allowing propagation of ARG uncertainty to downstream phylodynamic inferences; (3) explicit incorporation of spatial structure, enabling integrated pathogen genealogy–geography models; and (4) model extensions for recombination processes in viruses and bacteria that diverge from classical coalescent assumptions.

\paragraph{Cell lineage tracing with cell states}
At the cellular scale, the integration of lineage tracing with multi-omics data offers a powerful new frontier, which provides many functional insights on the developmental states and biological roles of sequenced cells. Jointly inferring lineage trees and cell states--encompassing both differentiation trajectories and spatial or migratory dynamics--will deepen our understanding of development, homeostasis, and cancer evolution (\fref{fig:fig3}C). This integrative direction will bridge the gap between ancestry (“where do cells come from?”) and function (“what do cells do?”) or fate (“what do they become?”).

\paragraph{}

Ultimately, the continued convergence of methods across these domains--phylogenetics, population genetics, and cell lineage biology--promises to refine our reconstruction of evolutionary history and enables new biological questions to be asked and answered through the language of trees and graphs.

\section{Acknowledgments}

We thank Rasmus Nielsen and Yun Song for helpful discussions, and Sebastian Prillo for helping formulate the symposium proposal and perspective. YD is supported by NIH grant R01HG014005. SHZ is supported by an NDPH Intermediate Fellowship (Oxford Population Health). YZ is supported by NSF CAREER award 2338710. CZ is supported by National Natural Science Foundation of China (NSFC). BC is supported by National Natural Science Foundation of China (Grant No.32300492); Major Project of Guangzhou National Laboratory (Grant No.GZNL2023A02006); Guangdong Provincial Natural Science Foundation (Grant No. 2025A1515011337); Project of State Key Laboratory of Respiratory Disease (Grant No. SKLRD-Z-202402)

\section{Conflicts of interest}

The authors declare no conflicts of interest.

\bibliographystyle{unsrtnat}
\bibliography{references}

\clearpage

\end{document}